\documentclass[12pt]{article}
\usepackage{cite}
\usepackage{amssymb}
\usepackage{graphicx}

\def\labelmark{}
\def\void{}

{\ifx\void\labelname\def\junk{\end{displaymath}}
\else\def\junk{\end{eqnarray}}\fi\junk\labelmark\def\labelname{}}

\begin{document}
\begin{titlepage}
\setcounter{page}{1}
\renewcommand{\thefootnote}{\fnsymbol{footnote}}

\begin{flushright}
hep-th/0309085
\end{flushright}

\vspace{6mm}
\begin{center}

{\Large\bf Anisotropic Quantum Hall Matrix Model}

\vspace{8mm} {\bf Ahmed Jellal$^1$\footnote{{\textsf
jellal@gursey.gov.tr }}}, 
{\bf Rudolf A. R\"omer$^2$\footnote{{\textsf r.roemer@warwick.ac.uk}}} 
and {\bf Michael
Schreiber$^1$\footnote{{\textsf schreiber@physik.tu-chemnitz.de}}}\\

\vspace{5mm} $^1${\em  Institut f\"{u}r Physik, Technische
Universit\"{a}t,\\ D-09107 Chemnitz, Germany }\\

$^2$ {\em Department of Physics, University of Warwick,\\
Coventry CV4 7AL, UK}\\

\end{center}

\vspace{5mm}
\begin{abstract}
We consider the anisotropic effect in the quantum Hall systems by
applying a confining potential that is not of parabolic type. This
can be done by extending Susskind--Polychronakos's approach to
involve the matrices of two coupled harmonic oscillators. Starting
from its action, we employ a unitary transformation to diagonalize
the model. The operators for building up the anisotropic ground
state and creating the collective excitations can be constructed
explicitly. Evaluating the area of the quantum Hall droplet,
we obtain the corresponding filling factor which is found to
depend on the anisotropy parameter
and to vary with the magnetic field strength. This can be used
to obtain the observed anisotropic filling factors, i.e.
${9\over 2}$, ${11\over 2}$ and others.

%
\end{abstract}

\end{titlepage}

\section{{Introduction}}

Recently it has been shown experimentally that the quantum Hall
(QH) phenomena can also appear in anisotropic systems. Indeed,
large transport anisotropies at half fillings $\nu={9\over 2},
{11\over 2}, \cdots$ in high-quality two-dimensional electron
gases (2DEG) in GaAs \cite{Lilly99,Du99} are seen. The central
observation is that the resistivity becomes strongly anisotropic
close to half filling of the topmost Landau level. From a
theoretical point of view, it is argued to be a signature of a
novel Coulomb-induced charge-density-wave ground state whose
existence had been predicted by Fogler {\it et al.}
\cite{Fogler96, Fogler01} and by Moessner and Chalker
\cite{Moessner96}. 

The aim of the present paper is to study the {\em anisotropic}
effect in the QH systems by means of the matrix-model language.
This can be done by extending Susskind--Polychronakos's (SP) idea
to coupled harmonic oscillators. In the SP approach it has been
suggested to replace the classical configuration space of $N$
electrons by a space of two $N\times N$ hermitian matrices and the
time component of the vector potential by a hermitian matrix. The
confining potential plays an important rule, since it defines the
Hamiltonian of the theory. Mainly, we are interested to
investigate the Laughlin liquid by considering a confining
potential that is not of parabolic type.

We develop an appropriate anisotropic model that generalizes the
SP approach and investigate the basic features of these QH fluids.
Making use of a unitary transformation, we end up with a
diagonalized system that allows us to define creation and
annihilation matrix operators. Calculating the area of the QH
droplet of matrix variables, we obtain a filling factor
$\nu_{\rm anis}$, which depends on an anisotropy parameter
$a(B)$, thus generalizing the isotropic (Polychronakos) factor
$\nu_{\rm p}$. We show that $\nu_{\rm anis}$ can be tuned to
describe some special anisotropic filling factors, i.e. ${9\over
2}$ and ${11\over 2}$. We build up the ground state of our model
as well as its excitations in terms of two different
representations, those corresponding to variables before and after
a suitable transformation.

In section \ref{sec-cmm}, we define our model by considering an
action, which involves a confining potential that is not of
parabolic type. Rotating the system by an angle $\varphi={\pi\over
2}$, we define its new action and determine the Gauss law
constraint as well as the equations of motion. Their solutions
will be given and these will be used to find the corresponding
solutions before the transformation in section \ref{sec-transform}. We
study our model quantum mechanically by deriving the Hamiltonian and
constructing a set of operators that lead to its quantization in
section \ref{sec-quantum}. Section \ref{sec-ground-state} is
devoted to building up the corresponding ground state as well as its
excitations and determining the filling factor. Finally we close
by emphasizing that under some conditions our ground state 
can be visualized similarly to the Laughlin
states with $\nu_{\rm p}={1\over k+1}$.

\section{{Coupled matrix model }}
\label{sec-cmm}

We start by recalling that Susskind \cite{susskind} recently
proposed an infinite non-commutative Chern-Simons matrix model for
describing the Laughlin liquid \cite{laughlin}. Subsequently,
Polychronakos \cite{polychronakos1} suggested a regularized
version of the Susskind model by introducing a bosonic field
$\psi$ that is a boundary term. Basically it is a {\rm finite}
non-commutative Chern-Simons matrix model and allows us to reproduce
the basic features of the Laughlin theory, i.e. the quantization
at the filling factor
\begin{equation}\label{polyff}
\nu_{\rm p}={1\over k+1}.
\end{equation}
The level $k$ of the Chern-Simons term is identified with
$B\theta$ by correspondence between the gauge fields and the
matrix variables at a large number $N$ of particles. $B$ and
$\theta$ are, respectively, magnetic field and non-commutativity
parameter.

In what follows, we investigate the anisotropic effect in the QH
systems by building up the ground state and determining its
filling factor. This can be done by generalizing the
SP action to the new action
\begin{equation}\label{TOT}
S ={B\over 2}\int dt \; {\rm{Tr}}\sum_{a,b=1}^{2}\left\{
\epsilon^{ab} \left(\dot{X}_a +i\left[A_0,X_a\right]\right) X_b +
2\theta A_0 -\omega X_a^2\right\} +{\mu}X_1 X_2 +
\psi^{\dag}\left(i \dot{\psi} -A_0 \psi\right)
\end{equation}
where $A_0$, $X_1$ and $X_2$ are classical 
hermitian-matrix-valued variables, 
$\epsilon^{ab}$ is the fully antisymmetric
tensor; ${\rm{Tr}}$ and $[\cdot,\cdot]$ denote operations in
matrix space. $\mu$ is playing the role of a coupling parameter
between two sectors parameterized by $X_1$ and $X_2$. We note
that~(\ref{TOT}) is an extension of two coupled harmonic
oscillators into matrix-model language. It is clear that by
switching off $\mu$, we recover the SP model. This suggests that
we are going to study a QH system of particles confined in an
anisotropic potential
\begin{equation}\label{OPOT}
V\left(X_1, X_2\right)= \omega \left(X_1^2+X_2^2\right)+ {\mu}X_1
X_2
\end{equation}
instead of a parabolic confinement with single frequency $\omega$.

As far as the action $S$ is concerned, the full symmetry is the
gauge group $U(N)$. The matrix-model variables transform under
this invariance as
\begin{equation}
\label{VTRAN} X_a \rightarrow U X_a U^{-1}, \quad \psi \rightarrow
U \psi.
\end{equation}
The gauge field $A_0$ ensures the gauge invariance of the action,
its equation of motion being the Gauss law constraint
\cite{polychronakos1}
\begin{equation}
\label{polg} {\cal{G}}\equiv [X_1,X_2] - i\theta\left( {{\bf 1}}-
{1\over B\theta}\psi \psi^{\dag} \right)=0 \quad
\end{equation}
representing the non-commutative aspects of the theory. The
equations of motion for the variables $X_1$ and $X_2$ are
different from the isotropic model. We find
\begin{equation}\label{XEOM}
\dot{X}_1 - \omega {X}_2 + {\mu\over B} {X}_1=0, \qquad \dot{X}_2
+ \omega {X}_1 - {\mu\over B} {X}_2=0 
\end{equation}
where the main difference with respect to the parabolic case is
the presence of the third term in both equations. These will be
solved in the next section after introducing a unitary
transformation. Essentially, their solutions will determine the
nature of the classical motion in the system.

\section{{Rotation into the principal-axis system}}
\label{sec-transform}
As~(\ref{TOT}) involves an interacting term,
for a straightforward investigation  of the basic
features of the system we
use an appropriate transformation. 
This can be done by defining new variables
\begin{equation}
\label{trans}
Y_a= N_{ab} X_b
\end{equation}
where the matrix 
\begin{equation}
\begin{array}{l} \label{mtrans}
 (N_{ab}) = \pmatrix{\cos{\varphi\over 2} & -\sin{\varphi\over 2}\cr
\sin{\varphi\over 2} &  \cos{\varphi\over 2}\cr}
\end{array}
\end{equation}
is the unitary rotation of the elliptic system
(\ref{OPOT}) into its principal-axis set with the mixing angle
$\varphi$. Clearly, $\varphi$ should satisfy the constraint
\begin{equation}
\varphi = \pi(n+{1/2}) 
\end{equation}
with $n$ an integer. Without loss of
generality we fix $n=0$. In this case, $S$ transforms as
\begin{equation}
\label{NAC} S' ={B\over 2}\int dt \; {\rm{Tr}}\sum_{a,b=1}^{2}\left\{
\epsilon^{ab} \left(\dot{Y}_a + i \left[A_0,Y_a\right]\right) Y_b
+ 2 \theta A_0 - \omega_a Y_a ^2 \right\} + \psi^{\dag}\left(i
\dot{\psi} - A_0 \psi\right)
\end{equation}
where the frequencies $\omega_1$ and $\omega_2$ are
\begin{equation}
\omega_1 = {\omega} +{\mu \over B}  ,\quad \omega_2 =
{\omega} -{\mu\over B}.
\end{equation}
Obviously $S'$ is also invariant under the gauge group $U(N)$ and
the matrix-model variables transform as~(\ref{VTRAN}). The
equations of motion of different matrix-model variables can be
derived in the usual way. It is easily seen that for $A_0$, we get
a Gauss law constraint analogous to 
(\ref{polg})
\begin{equation}
\label{gtran}
{\cal G}'\equiv [Y_{1}, Y_{2}] - i\theta \left( {\bf 1} - {1\over
B\theta}\psi \psi^{\dag} \right)=0
\end{equation}
where its trace part is given by
\begin{equation} \psi^{\dag}\psi
= N B\theta.
\end{equation}
For the variables $Y$ we obtain 
\begin{equation}\label{YEOM}
\dot{Y}_1 + {\omega_2\over B}{Y}_2 =0, \qquad
\dot{Y}_2 - {\omega_1\over B}{Y}_1 =0,
\end{equation}
similar to those derived by Polychronakos \cite{polychronakos1},
but involving different frequencies $\omega_1$ and $\omega_2$.
Their solutions can be written as
\begin{equation}\label{ysol}
{Y}_1 = {1\over 2a} \left(C e^{-i\Omega t}
+C^{\dag} e^{i\Omega t}\right), \qquad {Y}_2 = {i\over 2} a
\left(C e^{-i\Omega t} - C^{\dag} e^{i\Omega t}\right) 
\end{equation}
where the frequency is $\Omega =
\left(\omega_1\omega_2\right)^{1\over 2}$ and $a(B)>0$ is fixed
to be
\begin{equation}\label{alp}
a(B) = \left({B\omega+\mu\over
B\omega-\mu}\right)^{1\over 4}.
\end{equation}
We call it {\em anisotropy} parameter and note that
$B\omega>\mu$ must be
chosen. In Figure $1$, we show how $a(B)$ depends
on $B$ for some choices of $\omega$ and $\mu$. The $N\times N$
matrices $C$, $C^{\dag}$ appearing in the above equations are
defined by
\begin{equation}
C = a Y_1 + {i\over a}
Y_2, \qquad C^{\dag} = a Y_1 - {i\over a} Y_2.
\end{equation}
The equation of motion for the  bosonic field,
$(\dot{\psi}=0)$, can be solved to get \cite{polychronakos1}
\begin{equation}
\psi=\sqrt{NB\theta}| v\rangle
\end{equation}
where $| v \rangle$ is a constant vector of unit length. This can
be used to show that $C$ and $C^{\dag}$ satisfy the constraint
\begin{equation}
\left[C, C^{\dag}\right] = 2\theta \left( 1 - N |  v\rangle
\langle v | \right).
\end{equation}
The matrices $C$, $C^{\dag}$ determine the classical solution for
the action~(\ref{NAC}) which is given as \cite{polychronakos1}
\begin{equation}
\label{sysol}
(Y_1)_{mn} = y_n\delta_{mn}, \qquad (Y_2)_{mn} =
z_n\delta_{mn} - {i\theta \over y_m-y_n} (1-\delta_{mn}),
\end{equation}
solving the Gauss law constraint~(\ref{gtran}).

Now it is easily seen that the equations of motion~(\ref{XEOM})
can be solved once we express, at $\varphi={\pi\over 2}$,  $X_a$
in terms of the $Y_a$
\begin{equation}
{X}_1 =
{1\over \sqrt{2}} \left({Y}_1 + {Y}_2 \right), \qquad {X}_2 =
{1\over \sqrt{2}} \left({Y}_2 - {Y}_1 \right).
\end{equation}
Thus 
we end up with the classical solutions of two coupled 
harmonic oscillators.

\begin{figure}
\begin{center}
\includegraphics{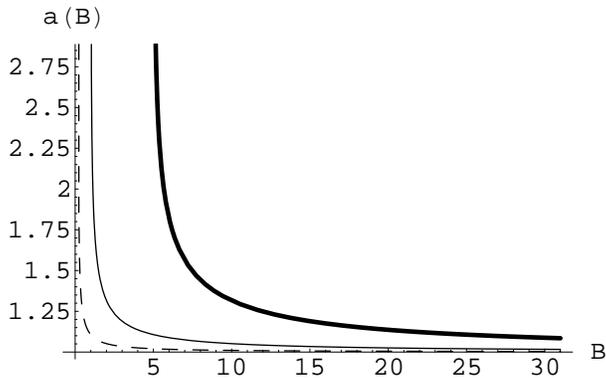}
\end{center}
\label{fig-alpha_B} \caption{Variation of $a(B)$ in terms of
magnetic field $B$. 
Dashed line: $\omega={\mu\over 5}$, {solid line}: $\omega=\mu$
and strong line: $\omega={\mu\over 5}$.}
\end{figure}

\section{Quantizing the theory}
\label{sec-quantum}

Upon quantization the matrix elements of $X_a, Y_a$ and the
components of the field $\psi$ become operators.  As usual, the
quantum Hamiltonian ${H}$ can be derived from the relation
\begin{equation}\label{HRLA}
{H} = {\dot X}{\partial{L}\over \partial{\dot X}} - L
\end{equation}
where ${\partial{L}\over \partial{\dot X}}$ defines the conjugate
momentum and $L$ is the Lagrangian corresponding to action
(\ref{NAC}). We find that ${H}$ can be written as a sum of a free
and an interacting part
\begin{equation}\label{HAMS}
 {H} = {\rm Tr} \left[{B\omega\over 2} \left(X_1^2 +X_2^2\right) - \mu
X_1X_2\right]
\end{equation}
which is nothing but the non-parabolic confining potential $V$.
This means that the kinetic energy is negligible compared to $V$.
With~(\ref{HAMS}), we actually have two possibilities to get the
Hamiltonian in terms of the matrices $Y$, either
by transforming ${H}$ via~(\ref{trans}) to obtain
\begin{equation}\label{HANP}
{H}' =\omega_2 \; {\rm Tr} \left( a^4 Y_1^2 + Y_2^2\right)
\end{equation}
or by starting straightforwardly from~(\ref{HRLA}), using
$S'$~(\ref{NAC}) to end up with~(\ref{HANP}). Let us remark
that~(\ref{HANP}) clearly shows  the anisotropy in ${H}'$,
i.e.\ the motion in the two coordinate directions $Y_{1,2}$ has
different frequencies. Fixing $\omega_1=\omega_2$, we recover the
Polychronakos Hamiltonian.

The form of ${H}'$ is similar to the harmonic oscillator and can
easily be diagonalized. We define two $N\times N$ matrices of
creation and annihilation operators
\begin{equation}
\label{creal} C_{nm} = \sqrt{B\over 2}\left[a (Y_1)_{nm} + 
{i\over a} (Y_2)_{nm}\right], \qquad C_{mn}^{\dag}
=\sqrt{B\over 2}\left[a (Y_1)_{nm} -  {i\over a}
(Y_2)_{nm}\right] .
\end{equation}
Their commutator can be evaluated by calculating from (\ref{NAC}) that
the operators $Y$ satisfy the commutation relation
\begin{equation} 
\left[ (Y_1) _{nm},~ (Y_2) _{n'm'} \right] = {i
\over B} \delta_{nm'} ~\delta _{n'm}.
\end{equation}
This implies
\begin{equation}
\label{heisa} \left[ C_{nm}, C_{n'm'}^{\dag}
\right]=\delta_{nm'}\delta_{n'm}
\end{equation}
where $n,m=1,\ldots, N$. Other commutators vanish. After some
algebra, we find that ${H}^{'}$ can be expressed in terms of $C$
and $C^{\dag}$ as
\begin{equation}
\label{ham1}
 {H}^{'}= \Omega'\;
\left( {\hat{N}}+ {N^{2}\over 2} \right)
\end{equation}
where $\Omega'={4\Omega / B}$ and the total number operator
\begin{equation}
{\hat{N}}= \sum_{n ,m=1}^{N} C_{mn}^{\dag} C_{nm}
\end{equation}
is counting the $N$ particles forming the system under
consideration.

\section{{Determining the ground states}}
\label{sec-ground-state}

Via the unitary transformation (\ref{mtrans}), we can also express
${H}$ in terms of the operators $C$, $C^{\dag}$ and build up the
anisotropic ground state of our model as well as its excitations
and determine the corresponding filling factor. First one has to
construct a physical state $|\Phi \rangle $ obeying the Gauss law
constraint~(\ref{polg}). To proceed, we determine $| \Phi'\rangle$
corresponding to ${H}'$ and go on to get $| \Phi \rangle$.

The ground state satisfying $ {\cal G}'| \Phi'\rangle =0$ can be
built as
\begin{equation}
\label{yvacu} | \Phi'\rangle = \bigl[ \epsilon ^{j_1...j_{N}}
\psi^{\dag}_{j_1} (\psi^{\dag}C^{\dag})_{j_2} \cdots (\psi^{\dag}
C^{\dag N-1})_{j_{N}} \bigr]^{k} | 0 \rangle
\end{equation}
where $\epsilon ^{j_1...j_{N}}$ are elements of the fully
antisymmetric tensor. Its energy spectrum is
\begin{equation} E_k =\Omega'\left[ {k\over 2}N(N-1)
+ {N^{2}\over 2}\right].
\end{equation}
Taking $\mu=0$, we recover the Hellerman--Van Raamsdonk ground
state \cite{hellerman1} constructed for the isotropic SP model.

Having the ground state $| \Phi' \rangle$, it is natural to ask
about its anisotropic fraction $\nu_{\rm anis}$. To answer this
question, we first calculate the area $A'$ of the QH droplet of
the matrices $Y$. This can be defined as
\begin{equation} \label{yarea}
A'= {\pi\over N} {\rm Tr} \left(Y_1^2 + Y_2^2\right).
\end{equation}
Mapping $A'$ in terms of the creation and annihilation matrix
operators and evaluating the trace with respect to (\ref{yvacu}),
we find in the large-$N$ limit
\begin{equation} \label{jel} \nu_{\rm anis} =
{\nu_{\rm p}\over 2}\left(a+{1\over a}\right).
\end{equation}
This relation is shown for some specific values of $\omega$
and $\mu$ in Figure $2$. In the limit $a=1$,
$\nu_{\rm anis}$ coincides with $\nu_{\rm p}$.

For a state that in the isotropic
system would correspond to the filling factor 
\begin{equation}
\nu_{\rm p}(B)=2\pi
\rho_0/B
\end{equation}
where $\rho_0$ is the density, an anisotropic confining
potential \cite{Lilly99,Du99} will yield an apparent filling
factor $\nu_{\rm anis}$ in dependence on $a(B)$,
i.e. on $B$ according to (\ref{jel}).
More precisely,
the filling factor ${9\over 2}$ can be obtained from (\ref{jel})
by considering the following configuration 
\begin{equation}
\nu = 1, \qquad
a ={9\over 2}\pm {\sqrt{77}\over 2}, 
\end{equation}
while for the factor
${11\over 2}$, one can choose 
\begin{equation}
\nu = 1, \qquad a ={11\over 2}\pm
{\sqrt{117}\over 2}. 
\end{equation}
%
More
generally, the anisotropic Hall conductivity is
\cite{Fogler01,halperin}
\begin{equation} \label{fog} \sigma_{\rm H}= {e^2\over
\hbar}\left(N+{1\over 2}\right) = {e^2\over \hbar} \nu_{\rm anis}
\end{equation}
where $N$ is the total number of particles. Comparing~(\ref{jel})
and~(\ref{fog}), we find
\begin{equation} a = {1\over
\nu_{\rm p}} \left(N+{1\over 2}\right) \pm {1\over \nu_{\rm p}}
\sqrt{\left(N+{1\over 2}\right)^2 -\nu_{\rm p}^2}.
\end{equation}

The excited states for the anisotropic model can be constructed as
\begin{equation}
\label{YEXIT} | \Phi'\rangle_{\rm ex} = \prod_{n=1}^{N-1} ({\rm
Tr}C^{\dag n})^{c_n} \bigl[ \epsilon ^{j_1...j_{N}}
\psi^{\dag}_{j_1} (\psi^{\dag}C^{\dag})_{j_2} \cdots (\psi^{\dag}
C^{\dag N-1})_{j_{N}} \bigr]^{k} ~ |0 \rangle
\end{equation}
and their energies are
\begin{equation}
E_k(c_n) =\Omega'\left[ {k\over 2}N(N-1) + {N^{2}\over 2}
+\sum_{n=1}^{N} n c_n\right]
\end{equation}
where the $c_n$ are non-negative integers. 

The anisotropic ground
state $| \Phi \rangle$ corresponding to the matrices $X$ can be
obtained by expressing the matrices $C$, $C^{\dag}$ in terms of
those corresponding to $X$. They are defined by
\begin{equation}
\label{MHOO} A_{nm} = \sqrt{B \over 2} \left(X_1 + i
X_2\right)_{nm}, \qquad A_{nm}^{\dag} = \sqrt{B \over 2} \left(X_1
- i X_2\right)_{nm}.
\end{equation}
By using the unitary transformation~(\ref{trans}),
(\ref{creal}) can be written as
\begin{equation}
\label{ACR} C= {\bar\eta} \left(A -i  A^{\dag}\right), \qquad
C^{\dag}= {{\eta}} \left(A^{\dag} +i  A\right)
\end{equation}
where $\eta$ is given by
\begin{equation}
 \eta = {1\over 2\sqrt{2}}\left\{ \left(a-{1\over a}\right) -
i \left(a + {1\over a}\right) \right\}
\end{equation} 
Inserting
(\ref{ACR}) into~(\ref{yvacu}) and normalizing, we obtain 
\begin{equation}
\label{xvacu} | \Phi \rangle = \left\{ \epsilon ^{j_1...j_{N}}
\psi^{\dag}_{j_1} \left[{\eta}\psi^{\dag}\left (A^{\dag} +
i A\right)\right]_{j_2} \cdots
\left[{\eta}^{N-1}\psi^{\dag}\left (A^{\dag} + i A
\right)^{N-1}\right]_{j_{N}}\right\}^k ~ | 0 \rangle.
\end{equation}
Its filling factor can be obtained from the
corresponding area
\begin{equation} A= {\pi\over N} {\rm Tr} (X_1^2+
X_2^2).
\end{equation}
Using the mapping (\ref{trans}), one can see that
$A$ coincides with $A'$ (\ref{yarea}), as
it should so that the ground
states~(\ref{xvacu}) and~(\ref{yvacu})
have the same fraction $\nu_{\rm anis}$.

The excited states of ${H}$ can be derived from~(\ref{YEXIT}) in
the same way as (\ref{xvacu}), giving
\begin{eqnarray}
\label{XEXIT}
\lefteqn{
|\Phi \rangle_{\rm ex} = {\prod_{n=1}^{N-1}}
\left[{\rm Tr} \; {\eta}^n
\left(A^{\dag} + i A\right)^{n}\right]^{c_n}}
\nonumber\\
& & \left\{ \epsilon ^{j_1...j_{N}}
\psi^{\dag}_{j_1} \left[{\eta}\psi^{\dag}
\left(A^{\dag} + i A\right)\right]_{j_2} 
\cdots 
\left[{\eta}^{N-1}\psi^{\dag}
\left(A^{\dag} + 
i A\right)^{N-1}\right]_{j_{N}}\right\}^k ~| 0 \rangle.
\end{eqnarray}
We close this section by noting that all the states obtained
here coincide with those for the isotropic SP model if we set
$a(B)=1$.

\begin{figure}
\begin{center}
\includegraphics{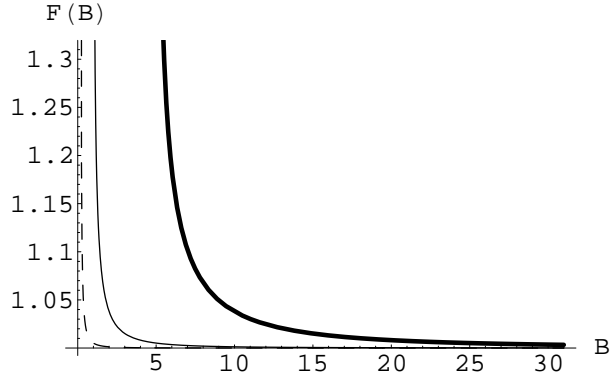}
\end{center}
\label{2} \caption{Variation of $F(B)={\nu_{\rm anis} /\nu_{\rm
p}}$ in terms of magnetic field $B$. 
Dashed line: $\omega={\mu\over 5}$, {solid line}: $\omega=\mu$
and strong line: $\omega={\mu\over 5}$.}
\end{figure}

\section{{Concluding remarks}}
\label{sec-concl}

We emphasize that the resulting filling factor 
for the anisotropic system can be mapped onto 
the isotropic one (\ref{polyff})
after making use of a second transformation \cite{polypriv}.
This can be
done by defining
the new matrices \begin{equation}
 W_1 = a Y_1,\qquad   W_2 = {1\over a} Y_2.
\end{equation} In terms of these, the system becomes identical to the
original rotationally symmetric matrix model proposed by
Polychronakos \cite{polychronakos1}. This has a circular droplet
ground state with the standard filling factor. The
transformation
\begin{equation}
 X_1,X_2 \longrightarrow Y_1,Y_2 \longrightarrow W_1,W_2
\end{equation}
is area-preserving, the first is a rotation by ${\pi\over 2}$ and
the second is a dilation by ''$a$'' in one direction and 
''${1\over a}$'' in the other. In this case, 
we end up with the filling factor (\ref{polyff}). 

In summary, we investigated the Laughlin liquids by considering a
confining potential that is of anisotropic type. This can be done
by applying matrix-model theory to
two coupled harmonic oscillators 
and studying the Laughlin states of the FQHE. Our model is
a generalization of that proposed by Susskind and Polychronakos
and reproduces its basic features in the limit $\mu=0$. Our most
important result shows that the filling factor is anisotropy
dependent. This suggests that our model may be a good starting
point for the interpretation of recent experiments in anisotropic
2DEG.

A direct connection of our model to the Calogero-Sutherland (CS)
models \cite{calogero,Sut85} as in Refs.\
\cite{polychronakos1,polychronakos2} would be highly beneficial.
An interpretation of the anisotropy of $X_1$ and $X_2$ in such a
mapping as two coupled CS models or as CS models with two types of
particles might be possible. While such models exist
\cite{SutR93,SutRS94}, they are solvable only for special sets of
parameters and the inclusion of a continuously varying $\mu$
appears quite challenging.

\section*{{Acknowledgments}}
We are very grateful to A.P.\ Polychronakos for helpful comments
and to the Deutsche Forschungsgemeinschaft, Schwerpunktprogramm
``Quantum-Hall-Effekt'', for financial support.


\begin{thebibliography}{1}\frenchspacing

\bibitem{Lilly99} {M.~P.} {Lilly},
 {K.~B.} {Cooper}, {J.~P.} {Eisenstein},
 {L.~N.} {Pfeiffer} and {K.~W.} {West},
 {Phys. Rev. Lett.} {82}, {394} (1999).

\bibitem{Du99}R.~R. Du, D.~C. Tsui,
H.~L. St\"ormer, L.~N. Pfeiffer and K.~W. West, Solid State Comm.
{109}, 389 (1999).


\bibitem{Fogler96} {M.~M.} {Fogler},
{A.~A.} {Koulakov} and {B.~I.} {Shklovskii}, {Phys. Rev. B.} {54},
{1853} (1996).

\bibitem{Fogler01}
{For a recent review, see M. M. Fogler, {\it Stripe and bubble
phases in quantum Hall systems}, (2001) cond-mat/0111001}.

\bibitem{Moessner96} {R.}~{Moessner} and {J.}~{Chalker},
{Phys. Rev. B} {54}, {5006} ({1996}).









\bibitem{susskind} L.~Susskind,
{\it The Quantum Hall Fluid and Non-Commutative Chern Simons
Theory}, (2001) {{hep-th/0101029}}.

\bibitem{laughlin} R.B.~Laughlin, Phys.\ Rev.\ Lett.
{ 50}, 1395 (1983).


\bibitem{polychronakos1} A.P. Polychronakos,
{JHEP} {0104}, 011 (2001).

\bibitem{hellerman1} S. Hellerman and  M. Van Raamsdonk,
{JHEP} {0110}, 039 (2001).

\bibitem{polychronakos2} A.P. Polychronakos,
JHEP {0106}, 070 (2001).












\bibitem{halperin} F. von Oppen, B.I. Halperin and A. Stern,
{\it Striped Quantum Hall Phases}, (2000) cond-mat/0002087.

\bibitem{polypriv} A.P.\ Polychronakos, private communication.

\bibitem{calogero}F. Calogero, J. Math. Phys. {12}, 419 (1971).
\bibitem{Sut85}
B. Sutherland,  in {\em Exactly Solvable Problems in Condensed
Matter and
  Relativistic Field Theory (Vol. 242)}, edited by B.~S. Shastry, V. Singh, and
  S.~S. Jha (Springer, New York, 1985).

\bibitem{SutR93}
B. Sutherland and R.~A. {R\"{o}mer}, Phys. Rev. Lett. {71}, 2789
(1993).

\bibitem{SutRS94}
B. Sutherland, R.~A. {R\"{o}mer}, and B.~S. Shastry, Phys. Rev.
Lett. {73}, 2154  (1994).
\end{thebibliography}

\end{document}